\documentclass[twocolumn,english,aps,prb,citeautoscript,superscriptaddress,altaffillsymbol]{revtex4-1}
\usepackage{lmodern}
\usepackage{lmodern}
\usepackage[T1]{fontenc}
\usepackage[latin9]{inputenc}
\usepackage{textcomp}
\usepackage{amsmath}
\usepackage{graphicx}

\makeatletter
\newcommand{\textgreek}[1]{$\beta$}

\usepackage{babel}

\usepackage{babel}

\usepackage{babel}

\makeatother

\usepackage{babel}
\begin{document}

\title{Quantum-optical spectroscopy of a two-level system using an electrically
driven micropillar laser as resonant excitation source}

\author{Sören~Kreinberg}

\affiliation{Institut für Festkörperphysik, Technische Universität Berlin, Germany}

\author{Tomislav~Grbeši\'{c}}

\affiliation{Institut für Festkörperphysik, Technische Universität Berlin, Germany}

\author{Max~Strauß}

\affiliation{Institut für Festkörperphysik, Technische Universität Berlin, Germany}

\author{Alexander~Carmele}

\affiliation{Institut für Theoretische Physik, Technische Universität Berlin,
Germany}

\author{Monika Emmerling}

\affiliation{Technische Physik, Julius-Maximilians-Universität Würzburg, Germany}

\author{Christian~Schneider}

\affiliation{Technische Physik, Julius-Maximilians-Universität Würzburg, Germany}

\author{Sven~Höfling}

\affiliation{Technische Physik, Julius-Maximilians-Universität Würzburg, Germany}
\affiliation{SUPA, School of Physics and Astronomy, University of St Andrews, St Andrews, KY16 9SS, United Kingdom}

\author{Xavier~Porte}

\affiliation{Institut für Festkörperphysik, Technische Universität Berlin, Germany}

\author{Stephan~Reitzenstein}
\email[Correspondence and requests for materials should be addressed to S.R., e-mail address: ]{stephan.reitzenstein@physik.tu-berlin.de}

\affiliation{Institut für Festkörperphysik, Technische Universität Berlin, Germany}
\begin{abstract}
Two-level emitters constitute main building blocks of photonic quantum
systems and are model systems for the exploration of quantum optics
in the solid state. Most interesting is the strict-resonant excitation
of such emitters to generate close to ideal quantum light and to control
their occupation coherently. Up till now related experiments have
been performed exclusively using bulky lasers which hinders the application
of resonantly driven two-level emitters in photonic quantum systems.
Here we perform quantum-optical spectroscopy of a two-level system
using a compact high-$\beta$ microlaser as excitation source. The
two-level system is based on a semiconductor quantum dot (QD), which
is excited resonantly by a fiber-coupled electrically driven micropillar
laser. In this way we dress the excitonic state of the QD under continuous
wave excitation and trigger the emission of single-photons with strong
multi-photon suppression ($g^{(2)}(0)=0.02$) and high photon indistinguishably
($V=57\pm9\%$) via pulsed resonant excitation at 156 MHz. 
\end{abstract}
\maketitle
The physics of two-level systems constitutes the basis for quantum
optics and quantum cavity electrodynamics (cQED). It has also important impact in the field of photonic quantum technologies
where it enables the secure exchange of information via single photons
\cite{Gisin.2007.,Bouwmeester.2010.,Nielsen.2010.,Ladd.2010.}. In
this context, semiconductor quantum dots (QDs) are close to ideal
two-level systems and can act as triggered sources of single photons
\cite{Aharonovich.2016.}. In order to explore their physics and the quantum nature of emission,
different excitation schemes have been developed which include simple
non-resonant electrical and optical excitation as well as more advanced
schemes like wetting-layer or p-shell resonant excitation \cite{Michler.2000.,Santori.2001.Phys.Rev.Lett.,Santori.2002,Ester.2008.,Heindel.2010,Gschrey.2015}. Most interesting is strict-resonant excitation of the fundamental
QD transition leading to resonance fluorescence \cite{Muller.2007.Phys.Rev.Lett.,Vamivakas.2009.NatPhys,He.2013}.
From an experimental point of view, strict-resonant excitation is very
demanding because it requires laser stray-light suppression by typically
more than 6 orders of magnitude \cite{Kuhlmann.2013.,Ates.2009,Hopfmann.2016.Semicond.Sci.Technol.}.
Nevertheless, the development of efficient suppression schemes and
the availability of mode-hop-free tunable lasers has led to huge progress
in this field and resonance fluorescence has become an important experimental
technique in quantum nanophotonics. Strict-resonant excitation has
for instance been used to study the subnatural linewidth from a single
quantum dot \cite{Matthiesen.2012} and to explore the non-resonant
dot-cavity coupling in microcavity systems \cite{Ates.2009.NatPhot3}.
It is interesting to note that up until now related experiments have
exclusively been performed using bulky and expensive tunable lasers.

In view of applications in quantum communication, strict-resonant
excitation of QDs is very advantageous because it leads to the emission
single photons with close to ideal quantum properties in terms of
multiphoton suppression and photon indistinguishably \cite{He.2013}.
Both aspects are crucial for advanced quantum communication protocols
based on entanglement distribution via Bell-state measurements \cite{Briegel.1998,Sangouard.2007.Phys.Rev.A}.
In addition, to enable ''real-world'' applications it is highly
interesting to develop compact electrically driven quantum light sources.
Unfortunately, standard excitation schemes based on carrier injection
via a pin-diode design are intrinsically non-resonant which limits
the achievable degree of indistinguishably \cite{Schlehahn.2016}.
To overcome this issue an advanced excitation concept has been developed
using an electrically driven microlaser to excite a single quantum
dot in a nearby microcavity system \cite{Munnelly.2017}. In this
concept, quasi-resonant p-shell excitation was shown \cite{Stock.2013.}
but strict-resonant excitation has not yet been achieved. In a similar
scheme a light emitting diode was used for on-chip excitation of a
single quantum dot \cite{Lee.2017}.

\begin{figure}[bh]
\includegraphics[width=1\columnwidth]{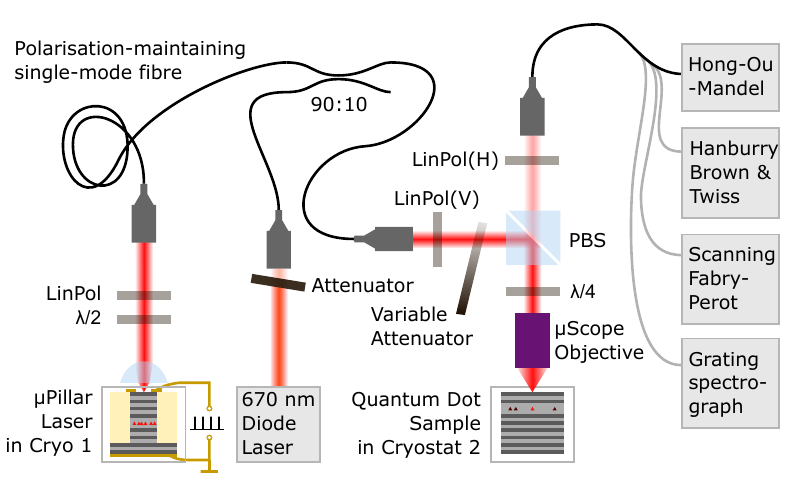} \caption{Schematic illustration of the experimental concept: Emission of the
electrically driven microlaser in cryostat 1 is fiber-coupled to resonantly
excite a single QD in cryostat 2. Applying either CW or pulsed excitation,
dressing of the two-level system or triggered emission of single photons
can be observed and verified by means of high-resolution spectroscopy
and single-photon counting.\label{figure1} }
\end{figure}

In this letter we demonstrate a fully nanophotonic approach to resonantly
drive a QD acting as two-level system and to generate single photons
with excellent multi-photon suppression and a high degree of photon
indistinguishably. Our concept is based on an electrically driven
quantum dot micropillar laser which resonantly drives a single QD
located in a planar microcavity. In order to resonantly excite a two-level
system we use a microlaser spectrally matched to a quantum dot,
where the temperature of the microlaser is used as fine-tuning knob
in resonance scans. The experiments are performed under continuous
wave (CW) and pulsed excitation of the electrically driven microlaser
to observe Mollow-triplet spectra and the triggered emission of single
photons with a Hong-Ou-Visibility of 60\%, respectively. Our results
show the potential of high-$\beta$ microlasers (in this case $\beta=1\%$,
see SI) to act as excitation sources in quantum optics experiments
and constitute an important step towards integrated quantum nanophotonic
circuits which rely on a small scale coherent light source for resonant
excitation of quantum emitters.

Our experimental concept is illustrated in Fig.~\ref{figure1}. It
includes a QD-micropillar laser which is located in cryostat 1 and
a spectrally matched QD in cryostat 2. The light emitted by the microlaser
is coupled into a 10\,m long polarization-maintaing single-mode fiber
which is connected to the input port of the resonance fluorescence
setup to excite the selected QD in cryostat 2. The microlaser is driven
by an electrical voltage supply capable of delivering an adjustable
DC bias and voltage pulses with a minimum width of 520\,ps and an
amplitude of up to 8\,volt at an repetition rate of up to 312.5\,MHz.
Resonance tuning with a tuning range of about 35\,GHz is enabled
by changing the temperature of the microlaser in the range of 64\,K
to 68\,K. The sample temperature of cryostat 2 is set to 7\,K to
minimize phonon induced decoherence \cite{Thoma.2016} and carrier
escape from the QDs in resonance fluorescence experiments. See Methods
section for details on the sample technology and on the experimental
setup.

For the planned quantum-optical studies it is crucial to couple emission
of the high-$\beta$ microlaser with sub-microwatt output power very
efficiently into a single-mode fiber connecting the two cryostats.
For this purpose we collimate the microlaser emission via a single
low-loss $f=20\thinspace\textrm{mm}$ aspheric lens in front of the
optical window of cryostat 1. In this context, we would like to note
that a slight deviation of the circular cross-section splits the fundamental 
transverse micropillar mode into two gain-coupled 
mode components with a spectral splitting of 11\,GHz. One of the two modes wins the 
gain competition and undergoes the lasing transition\cite{Leymann.2013,Redlich.2016}.
Emission of the lasing mode is selectively coupled to the polarization-maintaining
single-mode fiber via suitable orientation of a half-wave plate placed
in the collimated beam path and a subsequent beam-coupler.

\section*{Results}

\begin{figure}
\includegraphics[width=1\columnwidth]{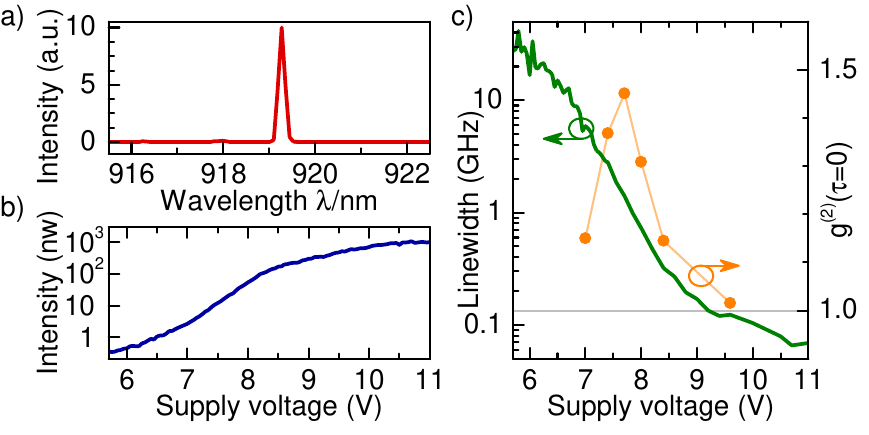}\caption{Characterization of the electrically driven micropillar laser under
CW excitation: \textbf{a}) \textmu EL-spectrum of the QD-micropillar
laser showing clean emission of the fundamental mode. Higher-order
lateral modes of the micropillar are well suppressed (see SI for details).
\textbf{b}) Input-Output dependence of the electrically driven QD-micropillar
laser with a threshold pump voltage of about 7\,-\,8\,V. \textbf{c})
Equal-time second-order photon autocorrelation function (as measured)
and spectral linewidth of the QD-microlaser (deconvoluted). The non-linear
input-output characteristics in conjunction with the narrowing of
emission linewidth by more than three orders of magnitude and the
transition of $g^{(2)}(0)$ from values larger than one to unity are
clear indications of predominantly stimulated emission of the QD-microlaser
above threshold. \label{figure2}}
\end{figure}

In the following we apply an electrically driven microlaser to demonstrate
for the first time the high potential of micro- and nanolasers in
quantum-optical spectroscopy. Indeed, while the research interest
on miniaturized lasers has increased rapidly in recent years, their
applicability as resonant excitation sources in quantum nanophotonics
has been widely unexplored up until now. To enable related studies
under strict-resonant excitation, it is crucial to find a microlaser
which a) can be operated electrically under CW and pulsed operation
with an emission pulse-width significantly shorter than the spontaneous
emission lifetime (here 510\,ps) of the QD, b) shows single-mode
emission with an emission linewidth significantly smaller than the
homogeneous linewidth of about 1 GHz, c) is spectrally matched with
a target QD within the available temperature-tuning range in the order
of 500\,GHz, and d) has sufficiently high optical output power of
about 100\,-\,500\,nW at the single-mode fiber output to at least
saturate the QD transition.

To meet these stringent requirements we first performed reference
measurements using a conventional tunable laser as excitation source
to select a QD showing pronounced and clean resonance fluorescence
at 920\,nm (see SI for more detail on the reference measurement),
where 920\,nm corresponds to the central wavelength reachable by
the micropillar lasers within the patterned array. In the second step
we chose a micropillar laser with slightly shorter emission wavelength
of 919\,nm at 10\,K, so that it can be spectrally matched with the
QD wavelength at 66\,K. Fig.~\ref{figure2}a shows the 32\,K \textmu EL
emission spectrum of the microlaser at the output of the single-mode
fiber. Without any spectral filtering we observe clear single-mode
emission with a side mode suppression ratio of 19\,dB and no significant
contribution from GaAs or wetting layer emission (see SI). Emission
of the laser is coupled into a single-mode fiber and leads to an output
power of 350\,nW at $V_{\textrm{bias}}=10.2\thinspace\textrm{V}$
at the fiber-output.

\begin{figure}
\includegraphics[width=0.5\columnwidth]{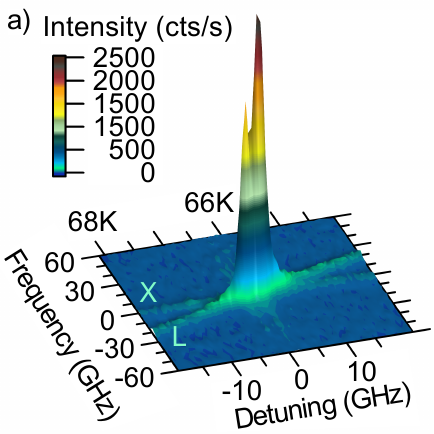}\includegraphics[width=0.5\columnwidth]{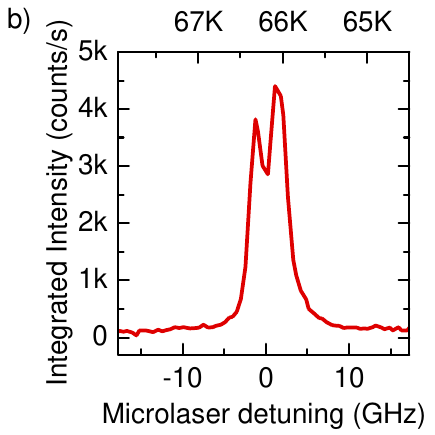}\caption{Resonance fluorescence (RF) of a single QD under CW microlaser-excitation
($V_{\textrm{bias}}=10.2\thinspace\textrm{V}$): \textbf{a}) 3D surface
plot of the QD emission intensity as a function of the frequency $f$
and the microlaser detuning $\Delta$. Using temperature-tuning in
a range of 64\,-\,68\,K the laser emission (L) is tuned through
the spectral resonance of the selected QD transition (X). Strong RF
signal is observed in resonance at about 66\,K. \textbf{b}) Emission
intensity of the QD vs. laser detuning integrated over the spectral
range displayed in \textbf{a} $-60\thinspace\mathrm{GHz}\leq f\leq60\thinspace\mathrm{GHz}$.
The double peak structure is attributed to the fine-structure splitting
of the excitonic transition.\label{figure3}}
\end{figure}

Figs. \ref{figure2}b and c present the corresponding voltage-dependent
output power and spectral linewidth of the micropillar laser, respectively.
The onset of laser action is indicated by the non-linear increase
of the output intensity between $V_{\textrm{bias}}$~=~7\,-\,8\,V
accompanied with a strong decrease of the emission linewidth to values
well below 0.1\,GHz. The associated transition from predominantly
spontaneous emission to stimulated emission is confirmed by measurements
of the bias voltage dependent second-order photon-autocorrelation
function $g^{(2)}(\tau)$, which shows the typical bunching behavior
in the threshold region with $g^{(2)}(0)>1$ and a transition towards
coherent emission associated with $g^{(2)}(0)=1$ at high excitation
\cite{Strauf.2006,Ulrich.2007}. It is worth noting that the equal-time
photon-correlation only approaches $g^{(2)}(0)=1$, when the linewidth
is already reduced by a factor of 100 in accordance with Ref. \cite{Kreinberg.2017}.

\section*{Discussion}

Having fulfilled all requirements a)-d) discussed above we are prepared
for resonance fluorescence (RF) experiments using the selected QD-micropillar
laser as coherent excitation source. For this purpose the temperature
of the fiber-coupled microlaser in cryostat 1 is gradually varied
between 64 K and 68 K and emission of the QD in the cyostat 2 is recorded
via the attached RF setup. The corresponding emission spectra (under
CW excitation) are presented Fig.~\ref{figure3}a as color-scale
intensity map. While under resonant conditions for QD-laser detuning
larger than 10\,GHz only weak emission of the QD and strongly suppressed
laser emission can be detected, strong and very pronounced RF emission
occurs at resonance. In fact, when scanning the microlaser emission
over the QD s-shell resonance, a double-peaked response with a splitting
of 5\,GHz detuning can be resolved, see Fig.~\ref{figure3}b. This
splitting is attributed to the fine-structure splitting of the excitonic
transition of the quantum dot\cite{Bayer.2002}. The measurements
presented in Figs. \ref{figure3}a and b also point out, that the
contribution of reflected laser light and the contribution of QD emission
due to above-band excitation by the red laser contributes marginally
to the RF signal.

\begin{figure}
\includegraphics[width=1\columnwidth]{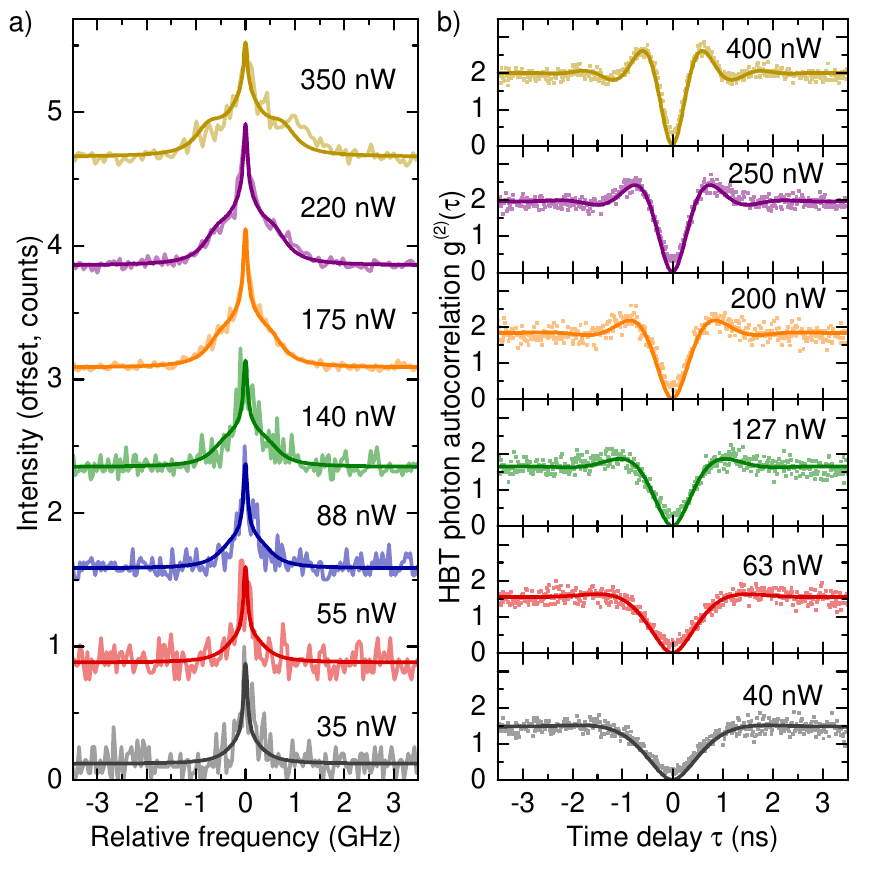}\caption{Excitation dependent resonance fluorescence (RF) emission spectra
and photon-autocorrelation function under CW microlaser-excitation:
\textbf{a}) High resolution RF emission spectra for different excitation
powers. With increasing excitation power we observe a transition of
a single emission line towards a Mollow-triplet like emission spectrum.
\textbf{b}) Second-order photon auto-correlation function $g^{(2)}(\tau)$
of the resonantly driven QD. The strong anti-bunching at zero time
delay $\tau=0$ indicates single-photon emission. At higher laser
powers, the narrowing of the anti-bunching peak together with the
directly visible Rabi oscillations indicate coherent excitation of
the two-level system. \label{figure4}}
\end{figure}

Coherent excitation of the QD based two-level system is demonstrated
in Fig.~\ref{figure4}. Changing the bias voltage of the micropillar
laser we cover a CW excitation power range of 35\,nW to 400\,nW.
Fig.~\ref{figure4}a shows the corresponding emission spectra recorded
with a high-resolution Fabry-Perot scanning interferometer. With increasing
excitation power we observe the characteristic line-broadening of
the single emission line with a measured FWHM of 600\,MHz at low
drive towards the evolution of the Mollow-triplet at high excitation
strength~\cite{Mollow.1969}. The splitting of the outer lines of
the Mollow-triplet with respect to the center amounts to 640\,MHz
at 350\,nW. The occurrence of this important signature of coherent
excitation is confirmed for the studied QD by reference measurements
over a wider range of excitation powers using a standard tunable laser
(see SI).

The quantum nature of RF emission is investigated by measuring the
second-order photon auto-correlation function $g^{(2)}(\tau)$, again
under CW excitation via the electrically driven microlaser. As can
be seen Fig.~\ref{figure4}b the excitation power dependent photon-correlation
reveals pronounced anti-bunching statistics with strong suppression
of multi-photon emission events associated with $g^{(2)}(0)<0.4$.
Upon increasing the excitation power from 40\,nW to 400\,nW the
simple antibunching dip evolves into a periodically modulated auto-correlation
function. The observed signatures are associated with Rabi-oscillations
in agreement with the Mollow-triplet observed in frequency domain
in Fig.~\ref{figure4}a. Noteworthy, we observe $g^{(2)}(\tau)>1$
in the vicinity of zero time delay $\tau\approx0$. This photon-bunching 
increases slightly with excitation power and indicates blinking of 
the QD due to metastable processes\cite{Davanco.2014.Phys.Rev.B}.

\begin{figure}
\includegraphics[width=1\columnwidth]{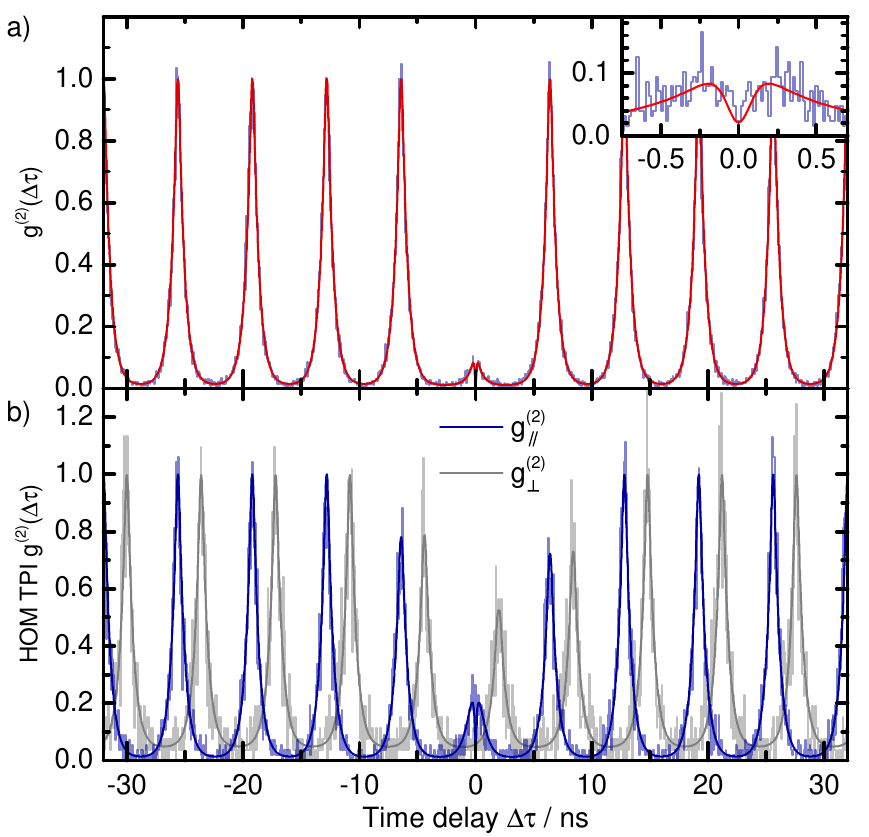}\caption{Demonstration of triggered single-photon emission and photon indistinguishably
under pulsed microlaser exitation: \textbf{a}) Second-order photon
auto-correlation under pulsed resonant excitation of the QD (pulse
area: $0.9\pi$). Triggered single-photon emission is clearly demonstrated
by strong anti-bunching with $g^{(2)}(0)\ll0.5$. The inset shows
that the non-ideal $g^{(2)}(0)$-value can mainly be attributed to
repeated QD excitation and decay within one single long-lasting laser
pulse. This interpretation is confirmed by numeric modelling (red
solid trace). \textbf{b}) HOM-histograms measured under co-polarized
and cross-polarized (shifted by $\delta\tau=2\thinspace\mathrm{ns}$
for the sake of clarity) configuration, respectively. Taking into
account the non-ideal $g^{(2)}(0)$-value of the data presented in
panel \textbf{a} we determine a HOM-visibility of $V_{\textrm{pure}}=0.57(9)$.
\label{figure5}}
\end{figure}

In order to obtain more detailed insight into the RF emission features
and to theoretically describe the experimental data presented in Fig.~\ref{figure4}
we consider the QD as two level system with a spontaneous emission
lifetime $T_{1}$, a dephasing time $T_{2}$, and an excitation power-dependent
Rabi frequency $\Omega$. The properties $T_{1}=510\thinspace\textrm{ps}$
and $\Omega/\sqrt{P}=2\pi\times1.33\thinspace\textrm{THz}\thinspace\textrm{W}^{-1/2}$
were determined via time resolved experiments under pulsed micropillar
laser excitation (cf. Fig. \ref{figure5}) and via excitation power
dependent auto-correlation investigations using a standard tunable
laser (see SI), respectively.

Using the formulas introduced in the SI we are able to model the experimental
data under variation of $T_{2}$. All optimum values of $T_{2}$ lie
in the vicinity of 500\,ps. Assuming $T_{2}$~=~500\,ps as given,
we obtain excellent quantitative agreement between experiment and
theory as can be seen in Fig.~\ref{figure4}a and Fig.~\ref{figure4}b,
where solid lines present the calculated data from formulas S3 and
S4, respectively.

With respect to applications in photonic quantum technology it is
crucial to demonstrate the triggered emission of single photons with
excellent quantum properties. For this purpose we biased the microlaser
with $V_{\textrm{bias}}=4.71\thinspace\textrm{V}$ below the onset
of lasing and superimposed voltage pulses with $V_{\textrm{pp}}=8\thinspace\textrm{V}$
with width of 520\,ps and a repetition period of 6.4\,ns. It is
interesting to note that due to the nonlinear input-output dependence
of the microlaser the resulting optical emission pulses were shortened
significantly to a width of 200\,ps (FWHM). The ratio of the peak
laser intensity to the strongest afterpulsing intensity was larger
than 18\,dB (see SI). Pulsed emission was again coupled via the single-mode
fiber and the RF configuration into cryostat 2 to resonantly excite
the selected QD. The corresponding photon autocorrelation function
was recorded at an excitation power of 22\,nW and is presented in
Fig.~\ref{figure5}a. Pulsed emission of light is clearly identified
by the train of correlation pulses separated by 6.4 ns, and triggered
single-photon emission is evidenced by the strongly reduced peak at
zero delay with $g^{(2)}(0)=2\%$. The zoom-in presentation $g^{(2)}(\tau)$
in the inset of Fig.~\ref{figure5}a shows a characteristic substructure
of the central $g^{(2)}(\tau)$ peak with a minimum at $\tau=0$ and
side-peaks at finite delay. This correlation feature indicates that
the non-ideal multi-photon suppression is mainly due to repeated QD
excitation and decay within one single long-lasting laser pulse. Numerical
modelling (red solid trace, cf. SI for details) was used to confirm
the nature of the central correlation feature and to extract the lifetime
of $T_{1}=510\thinspace\textrm{ps}$ and the pulse area of $0.9\pi$
by fitting the modelled curve to the experimental data. Indeed, it
was shown that both increased pulse length\cite{Fischer.2016} and
increased pulse area\cite{Fischer.2017}, increase the probability
of multiphoton-photon events. Thus, even better multi-photon suppression
may be achieved by applying shorter electrical pulses to the microlaser
in the future.

Finally, we study the photon-indistinguishably of emission under pulsed
microlaser-excitation via a fiber-coupled Hong-Ou-Mandel two-photon
interferometer with adjustible delay\cite{Bennett.2008,Thoma.2016}.
Here, the delay of the associated Mach-Zehnder interferometer was
matched to the pulse repetition rate of 6.4 ns of the electrical voltage
source used to drive the microlaser. The resulting photon-correlation
diagram of emission from the resonantly excited QD is displayed in
Fig.~\ref{figure5}b both in co-polarized and cross-polarized measurement
configuration. The experimental data is displayed as light blue and
light gray lines, fitted numerically modelled data (see SI) is displayed
as dark blue and dark grey lines. The only fitting parameter (except
for background counts and scaling) is the imbalance of the second
beam splitter, the one at which the HOM effect happens, turning out
to be 8:9 and resulting in different height of the peaks at $\pm6.4\thinspace\mathrm{ns}$.
A significant degree of photon indistinguishability is evidenced by
strongly reduced coincidences in the co-polarized case, while for
the cross-polarized case we observe $g_{\bot}^{(2)}(0)\approx0.5$
as expected for distinguishable photons. In order to determine the
resulting two-photon interference visibility $V$, we first integrate
areas $A_{n}^{\Vert,\bot}$ of the peaks centered at time delays $\tau=n\times6.4\textrm{\thinspace ns},\,n\in\left\{ -7,-6,-5,\ldots,6,7\right\} $
for each polarization configuration. Then, following 
\begin{align}
A_{\textrm{ref}}^{\Vert,\bot} & =\frac{1}{12}\sum_{n=2}^{7}A_{n}^{\Vert,\bot}+A_{-n}^{\Vert,\bot}\nonumber \\
V & =1-\frac{A_{0}^{\Vert}A_{\textrm{ref}}^{\bot}}{A_{\textrm{ref}}^{\Vert}A_{0}^{\bot}}\label{eq:visibility}
\end{align}
we extract a raw two-photon interference visibility of $V=0.44(4)$.
When compensating for the non-zero $g^{(2)}(0)$ and for slight HOM
beam-splitter imbalance of 8:9, we obtain a two-photon visibility
as high as $V_{\textrm{pure}}=0.57(9)$ (see SI for details). This
value is higher than 41\% reported in Ref.~\cite{Schlehahn.2016}
for direct non-resonant electrical excitation of a QD via carrier
injection in a pin-diode design. It is, however, significantly lower
than values exceeding 90\% achieved by resonant-excitation via standard
mode-locked lasers with ps-pulse widths. In order to explain the non-ideal
degree of photon-indistinguishably one can consider several possible
reasons such as temperature induced dephasing~\cite{Varoutsis.2005.Phys.Rev.B,Thoma.2016},
spectral fluctuations~\cite{Thoma.2016}. In the present case, i.e.
under resonant excitation at low temperature, we can exclude these
effects. Instead, we attribute the reduced HOM-visibility mainly to
the rather long optical pulse width of 200\,ps and to the non-Fourier-limited
dephasing time $T_{2}=0.5\textrm{\thinspace ns}\approx T_{1}<2T_{1}$
(see Fig. S9). The increased laser pulse-width in combination with
strong pulse-power ($0.9\pi$) leads to two-photon fluorescence pulses
and in turn results in reduced HOM-visibility \cite{Dada.2016,Fischer.2016,Fischer.2017}.
On the other hand, non-Fourier-limited $T_{2}$ directly makes photons
more distinguishable: Be it due to random phase changes or be it due
to finestructure splitting\cite{Santori.2002,Gazzano.2013} of the
QD transition and the thereby implied wavelength distinguishability.
We therefore expect a strong improvement of the photon-indistuinguishability
by carefully adjusting the detected polarization to one single QD
transition only and by reducing the optical pulse-length in future
studies.

\section*{Methods}

\textbf{Sample Technology} The QD-microlaser and the resonantly excited
QD are based on AlGaAs heterostructures grown by molecular beam epitaxy.
Both structures consist of high-quality AlAs/GaAs based on distributed
Bragg-reflectors (DBRs) forming a planar microcavity with a central
one-$\lambda$ GaAs cavity. A single layer of InGaAs quantum dots
acts as active medium. In case of the microlaser, the planar microcavity
is composed by a rather high number of 27 and 23 mirror pairs in the
n-doped lower and p-doped upper DBR to ensure pronounced light-matter
interaction and high-$\beta$ lasing. A dense array of micropillar
lasers with a diameter of 3\,\textmu m and a pitch of 60\,\textmu m
are realized by high-resolution electron-beam lithography and subsequent
reactive-ion etching. The sample is planarized with benzocyclobutene
(BCB) to mechanically support the ring-shaped upper Au-contacts, with
the positive side-effect of protecting the AlAs layers from oxidizing.
The realized array includes 62 electrically micropillar lasers emitting
in the spectral range 912\,-\,919\,nm. We refer to Ref.~\cite{Bockler.2008}
for further details on the fabrication of electrically contacted micropillars.
The QD sample used for the resonance fluorescence experiments has
a more asymmetric microcavity design with 24 and 5 mirror pairs in
the lower and upper DBR which fosters directional outcoupling of light
with an extraction efficiency of up to 42\% \cite{Maier.2014.}. Due
to the low QD density of $2\times10^{9}\thinspace\mathrm{cm}^{-2}$
and the presence of random photonic defects this planar microcavity
sample is very suitable of single QD experiments without the need
of lateral device processing.

\textbf{Experimental setup} The experimental configuration consits
of two optical tables, each of which includes a Helium-flow cryostat
(cryostat 1 and cryostat 2, respectively). The electrically driven
QD-microlaser is installed in cryostat 1 and a single-mode fiber guides
the laser light to the resonance fluorescence configuration in cross
polarization configuration \cite{Kuhlmann.2013.,Unsleber.2015,Strau.2016}
at cryostat 2. The resonant laser light enters the RF setup via a
fiber beamsplitter where it is superimposed with a low power non-resonant
support laser. The latter is a red diode laser (emission wavelength:
670\,nm) whose emission fills charge traps adjacent to the quantum
dot to effectively gate RF signal of the QD\cite{Nguyen.2012}. The
combined lasers are collimated again to free space and are aligned
with the RF detection beam path by means of a polarizing beam splitter
cube (PBS). Excitation of the quantum dot and detection of RF is then
performed confocally through a $N\negthinspace A=0.65$, $f=4\thinspace\textrm{mm}$
microscope objective. The main purpose of the PBS is to strongly suppresses
laser light reflected from the sample. In order to compensate for
possible polarization ellipticity and to maximize laser stray-light
suppression, a quarter-wave plate is placed in the excitation/detection
path beetween PBS and microscope objective.\cite{Strau.2016} The
detected light is fed into a polarization-maintaining single-mode
fiber, both for spatial filtering and to facilitate quantum-optics
experiments. For HBT- and HOM-style single-photon correlation experiments,
superconducting single-photon detectors (SSPDs) with a time resolution
FWHM of 55\,ps are correlated. High resolution RF spectra are recorded
by means of a scanning Fabry-Perot interferometer with 100\,MHz spectral
resolution.

\section*{Conclusion}

In summary we demonstrated a fully nanophotonic concept to control
single-photon emission of a solid-state two-level system. The concept
applies an electrically driven high-$\beta$ microlaser which resonantly
drives a semiconductor QD acting as two-level system. It shows for
the first time applicability of micro- and nano-laser in advanced
quantum-optics experiments under strict resonant excitation. Temperature
induced spectral fine-tuning of a suitable microlaser-QD allows us
to observe dressing of the fundamental QD transition and the occurrence
of Rabi-oscillations in photon-correlation measurements. Pulsed electrical
excitation of the microlaser leads the emission of single photons
with high multi-photon suppression ($g^{(2)}(0)=2\%$) and a Hong-Ou-Mandel
visibility as high as 57\%. As such our results present the great
potential of combining and coupling nanophotonic devices to systems
with enhanced functionality. In the future, our concept could be further
developed into a fully integrated on-chip resonantly pumped quantum
light sources with many interesting applications in photonic quantum
information technology.

\section*{Acknowledgements}

The research leading to these results has received funding from from
the European Research Council (ERC) under the European Union's Seventh
Framework ERC Grant Agreement No. 615613, from the German Research
Foundation via CRC 787 and Projects No. RE2974/5-1, RE2974/9-1 and SCHN1376/2-1,
the State of Bavaria and the German Ministry of Education and Research (BMBF) 
within Q.com-H.

\section*{Author contributions}

S.R. initiated the research and conceived the experiments. S.K. and
T.G. performed the experiments. X.P. and S.R. supervised the experiments.
A.C. did the continuous wave theoretical analysis, S.K. did the pulsed
theoretical analysis. M.S. built the resonance fluorescence setup.
M.E., C.S. and S.H. realized the samples. S.R. and S.K. wrote the
manuscript with contributions from all other authors.


\begin{thebibliography}{10}
\expandafter\ifx\csname url\endcsname\relax
  \def\url#1{\texttt{#1}}\fi
\expandafter\ifx\csname urlprefix\endcsname\relax\def\urlprefix{URL }\fi
\providecommand{\bibinfo}[2]{#2}
\providecommand{\eprint}[2][]{\url{#2}}

\bibitem{Gisin.2007.}
\bibinfo{author}{Gisin, N.} \& \bibinfo{author}{Thew, R.}
\newblock \bibinfo{title}{{Quantum communication}}.
\newblock \emph{\bibinfo{journal}{Nat. Photonics}}
  \textbf{\bibinfo{volume}{1}}, \bibinfo{pages}{165--171}
  (\bibinfo{year}{2007}).

\bibitem{Bouwmeester.2010.}
\bibinfo{author}{Bouwmeester, D.}, \bibinfo{author}{Ekert, A.~K.} \&
  \bibinfo{author}{Zeilinger, A.}
\newblock \emph{\bibinfo{title}{{The physics of quantum information: Quantum
  cryptography, quantum teleportation, quantum computation}}}
  (\bibinfo{publisher}{Springer}, \bibinfo{address}{Berlin},
  \bibinfo{year}{2010}).

\bibitem{Nielsen.2010.}
\bibinfo{author}{Nielsen, M.~A.} \& \bibinfo{author}{Chuang, I.~L.}
\newblock \emph{\bibinfo{title}{{Quantum Computation and Quantum Information}}}
  (\bibinfo{publisher}{{Cambridge University Press}},
  \bibinfo{address}{Cambridge}, \bibinfo{year}{2010}).

\bibitem{Ladd.2010.}
\bibinfo{author}{Ladd, T.~D.} \emph{et~al.}
\newblock \bibinfo{title}{{Quantum computers}}.
\newblock \emph{\bibinfo{journal}{Nature}} \textbf{\bibinfo{volume}{464}},
  \bibinfo{pages}{45--53} (\bibinfo{year}{2010}).

\bibitem{Aharonovich.2016.}
\bibinfo{author}{Aharonovich, I.}, \bibinfo{author}{Englund, D.} \&
  \bibinfo{author}{Toth, M.}
\newblock \bibinfo{title}{{Solid-state single-photon emitters}}.
\newblock \emph{\bibinfo{journal}{Nat. Photonics}}
  \textbf{\bibinfo{volume}{10}}, \bibinfo{pages}{631--641}
  (\bibinfo{year}{2016}).

\bibitem{Michler.2000.}
\bibinfo{author}{Michler, P.} \emph{et~al.}
\newblock \bibinfo{title}{{A quantum dot single-photon turnstile device}}.
\newblock \emph{\bibinfo{journal}{Science}} \textbf{\bibinfo{volume}{290}},
  \bibinfo{pages}{2282--2285} (\bibinfo{year}{2000}).

\bibitem{Santori.2001.Phys.Rev.Lett.}
\bibinfo{author}{Santori, C.}, \bibinfo{author}{Pelton, M.},
  \bibinfo{author}{Solomon, G.}, \bibinfo{author}{Dale, Y.} \&
  \bibinfo{author}{Yamamoto, Y.}
\newblock \bibinfo{title}{{Triggered single photons from a quantum dot}}.
\newblock \emph{\bibinfo{journal}{Phys. Rev. Lett.}}
  \textbf{\bibinfo{volume}{86}}, \bibinfo{pages}{1502--1505}
  (\bibinfo{year}{2001}).

\bibitem{Santori.2002}
\bibinfo{author}{Santori, C.}, \bibinfo{author}{Fattal, D.},
  \bibinfo{author}{Vu{\v{c}}kovi{\'c}, J.}, \bibinfo{author}{Solomon, G.~S.} \&
  \bibinfo{author}{Yamamoto, Y.}
\newblock \bibinfo{title}{{Indistinguishable photons from a single-photon
  device}}.
\newblock \emph{\bibinfo{journal}{Nature}} \textbf{\bibinfo{volume}{419}},
  \bibinfo{pages}{594--597} (\bibinfo{year}{2002}).

\bibitem{Ester.2008.}
\bibinfo{author}{Ester, P.} \emph{et~al.}
\newblock \bibinfo{title}{{p-Shell Rabi-flopping and single photon emission in
  an InGaAs/GaAs quantum dot}}.
\newblock \emph{\bibinfo{journal}{Physica E}} \textbf{\bibinfo{volume}{40}},
  \bibinfo{pages}{2004--2006} (\bibinfo{year}{2008}).

\bibitem{Heindel.2010}
\bibinfo{author}{Heindel, T.} \emph{et~al.}
\newblock \bibinfo{title}{{Electrically driven quantum dot-micropillar single
  photon source with 34{\%} overall efficiency}}.
\newblock \emph{\bibinfo{journal}{Appl. Phys. Lett.}}
  \textbf{\bibinfo{volume}{96}}, \bibinfo{pages}{10--13}
  (\bibinfo{year}{2010}).

\bibitem{Gschrey.2015}
\bibinfo{author}{Gschrey, M.} \emph{et~al.}
\newblock \bibinfo{title}{{Highly indistinguishable photons from deterministic
  quantum-dot microlenses utilizing three-dimensional in situ electron-beam
  lithography}}.
\newblock \emph{\bibinfo{journal}{Nat. Commun.}} \textbf{\bibinfo{volume}{6}},
  \bibinfo{pages}{7662} (\bibinfo{year}{2015}).

\bibitem{Muller.2007.Phys.Rev.Lett.}
\bibinfo{author}{Muller, A.} \emph{et~al.}
\newblock \bibinfo{title}{{Resonance fluorescence from a coherently driven
  semiconductor quantum dot in a cavity}}.
\newblock \emph{\bibinfo{journal}{Phys. Rev. Lett.}}
  \textbf{\bibinfo{volume}{99}}, \bibinfo{pages}{187402}
  (\bibinfo{year}{2007}).

\bibitem{Vamivakas.2009.NatPhys}
\bibinfo{author}{Vamivakas, A.~N.}, \bibinfo{author}{Zhao, Y.},
  \bibinfo{author}{Lu, C.-Y.} \& \bibinfo{author}{Atat{\"u}re, M.}
\newblock \bibinfo{title}{{Spin-resolved quantum-dot resonance fluorescence}}.
\newblock \emph{\bibinfo{journal}{Nat. Phys.}} \textbf{\bibinfo{volume}{5}},
  \bibinfo{pages}{198--202} (\bibinfo{year}{2009}).

\bibitem{He.2013}
\bibinfo{author}{He, Y.-M.} \emph{et~al.}
\newblock \bibinfo{title}{{On-demand semiconductor single-photon source with
  near-unity indistinguishability}}.
\newblock \emph{\bibinfo{journal}{Nat. Nano.}} \bibinfo{pages}{1--11}
  (\bibinfo{year}{2013}).

\bibitem{Kuhlmann.2013.}
\bibinfo{author}{Kuhlmann, A.~V.} \emph{et~al.}
\newblock \bibinfo{title}{{A dark-field microscope for background-free
  detection of resonance fluorescence from single semiconductor quantum dots
  operating in a set-and-forget mode}}.
\newblock \emph{\bibinfo{journal}{Rev. Sci. Instrum.}}
  \textbf{\bibinfo{volume}{84}}, \bibinfo{pages}{073905}
  (\bibinfo{year}{2013}).

\bibitem{Ates.2009}
\bibinfo{author}{Ates, S.} \emph{et~al.}
\newblock \bibinfo{title}{{Post-selected indistinguishable photons from the
  resonance fluorescence of a single quantum dot in a microcavity}}.
\newblock \emph{\bibinfo{journal}{Phys. Rev. Lett.}}
  \textbf{\bibinfo{volume}{103}}, \bibinfo{pages}{167402}
  (\bibinfo{year}{2009}).

\bibitem{Hopfmann.2016.Semicond.Sci.Technol.}
\bibinfo{author}{Hopfmann, C.} \emph{et~al.}
\newblock \bibinfo{title}{{Efficient stray-light suppression for resonance
  fluorescence in quantum dot micropillars using self-aligned metal
  apertures}}.
\newblock \emph{\bibinfo{journal}{Semicond. Sci. Technol.}}
  \textbf{\bibinfo{volume}{31}}, \bibinfo{pages}{095007}
  (\bibinfo{year}{2016}).

\bibitem{Matthiesen.2012}
\bibinfo{author}{Matthiesen, C.}, \bibinfo{author}{Vamivakas, A.~N.} \&
  \bibinfo{author}{Atat{\"u}re, M.}
\newblock \bibinfo{title}{{Subnatural linewidth single photons from a quantum
  dot}}.
\newblock \emph{\bibinfo{journal}{Phys. Rev. Lett.}}
  \textbf{\bibinfo{volume}{108}}, \bibinfo{pages}{093602}
  (\bibinfo{year}{2012}).

\bibitem{Ates.2009.NatPhot3}
\bibinfo{author}{Ates, S.} \emph{et~al.}
\newblock \bibinfo{title}{{Non-resonant dot--cavity coupling and its potential
  for resonant single-quantum-dot spectroscopy}}.
\newblock \emph{\bibinfo{journal}{Nat. Photonics}}
  \textbf{\bibinfo{volume}{3}}, \bibinfo{pages}{724--728}
  (\bibinfo{year}{2009}).

\bibitem{Briegel.1998}
\bibinfo{author}{Briegel, H.-J.}, \bibinfo{author}{D{\"u}r, W.},
  \bibinfo{author}{Cirac, J.~I.} \& \bibinfo{author}{Zoller, P.}
\newblock \bibinfo{title}{{Quantum Repeaters: The Role of Imperfect Local
  Operations in Quantum Communication}}.
\newblock \emph{\bibinfo{journal}{Phys. Rev. Lett.}}
  \textbf{\bibinfo{volume}{81}}, \bibinfo{pages}{5932--5935}
  (\bibinfo{year}{1998}).

\bibitem{Sangouard.2007.Phys.Rev.A}
\bibinfo{author}{Sangouard, N.} \emph{et~al.}
\newblock \bibinfo{title}{{Long-distance entanglement distribution with
  single-photon sources}}.
\newblock \emph{\bibinfo{journal}{Phys. Rev. A}} \textbf{\bibinfo{volume}{76}},
  \bibinfo{pages}{253} (\bibinfo{year}{2007}).

\bibitem{Schlehahn.2016}
\bibinfo{author}{Schlehahn, A.} \emph{et~al.}
\newblock \bibinfo{title}{{An electrically driven cavity-enhanced source of
  indistinguishable photons with 61{\%} overall efficiency}}.
\newblock \emph{\bibinfo{journal}{APL Photonics}} \textbf{\bibinfo{volume}{1}},
  \bibinfo{pages}{011301} (\bibinfo{year}{2016}).

\bibitem{Munnelly.2017}
\bibinfo{author}{Munnelly, P.} \emph{et~al.}
\newblock \bibinfo{title}{{Electrically Tunable Single-Photon Source Triggered
  by a Monolithically Integrated Quantum Dot Microlaser}}.
\newblock \emph{\bibinfo{journal}{ACS Photonics}} \textbf{\bibinfo{volume}{4}},
  \bibinfo{pages}{790--794} (\bibinfo{year}{2017}).

\bibitem{Stock.2013.}
\bibinfo{author}{Stock, E.} \emph{et~al.}
\newblock \bibinfo{title}{{On-chip quantum optics with quantum dot
  microcavities}}.
\newblock \emph{\bibinfo{journal}{Adv. Mater.}} \textbf{\bibinfo{volume}{25}},
  \bibinfo{pages}{707--710} (\bibinfo{year}{2013}).

\bibitem{Lee.2017}
\bibinfo{author}{Lee, J.~P.} \emph{et~al.}
\newblock \bibinfo{title}{{Electrically driven and electrically tunable quantum
  light sources}}.
\newblock \emph{\bibinfo{journal}{Appl. Phys. Lett.}}
  \textbf{\bibinfo{volume}{110}}, \bibinfo{pages}{071102}
  (\bibinfo{year}{2017}).

\bibitem{Thoma.2016}
\bibinfo{author}{Thoma, A.} \emph{et~al.}
\newblock \bibinfo{title}{{Exploring Dephasing of a Solid-State Quantum Emitter
  via Time- and Temperature-Dependent Hong-Ou-Mandel Experiments}}.
\newblock \emph{\bibinfo{journal}{Phys. Rev. Lett.}}
  \textbf{\bibinfo{volume}{116}}, \bibinfo{pages}{033601}
  (\bibinfo{year}{2016}).

\bibitem{Leymann.2013}
\bibinfo{author}{Leymann, H. A.~M.} \emph{et~al.}
\newblock \bibinfo{title}{{Intensity fluctuations in bimodal micropillar lasers
  enhanced by quantum-dot gain competition}}.
\newblock \emph{\bibinfo{journal}{Phys. Rev. A}} \textbf{\bibinfo{volume}{87}}
  (\bibinfo{year}{2013}).

\bibitem{Redlich.2016}
\bibinfo{author}{Redlich, C.} \emph{et~al.}
\newblock \bibinfo{title}{{Mode-switching induced super-thermal bunching in
  quantum-dot microlasers}}.
\newblock \emph{\bibinfo{journal}{New J. Phys.}} \textbf{\bibinfo{volume}{18}},
  \bibinfo{pages}{063011} (\bibinfo{year}{2016}).

\bibitem{Strauf.2006}
\bibinfo{author}{Strauf, S.}
\newblock \bibinfo{title}{{Self-Tuned Quantum Dot Gain in Photonic Crystal
  Lasers}}.
\newblock \emph{\bibinfo{journal}{Phys. Rev. Lett.}}
  \textbf{\bibinfo{volume}{96}}, \bibinfo{pages}{127404}
  (\bibinfo{year}{2006}).

\bibitem{Ulrich.2007}
\bibinfo{author}{Ulrich, S.~M.} \emph{et~al.}
\newblock \bibinfo{title}{{Photon statistics of semiconductor microcavity
  lasers}}.
\newblock \emph{\bibinfo{journal}{Phys. Rev. Lett.}}
  \textbf{\bibinfo{volume}{98}}, \bibinfo{pages}{043906}
  (\bibinfo{year}{2007}).

\bibitem{Kreinberg.2017}
\bibinfo{author}{Kreinberg, S.} \emph{et~al.}
\newblock \bibinfo{title}{{Emission from quantum-dot high-\textgreek{b}
  microcavities: Transition from spontaneous emission to lasing and the effects
  of superradiant emitter coupling}}.
\newblock \emph{\bibinfo{journal}{Light Sci. Appl.}}
  \textbf{\bibinfo{volume}{6}}, \bibinfo{pages}{e17030} (\bibinfo{year}{2017}).

\bibitem{Bayer.2002}
\bibinfo{author}{Bayer, M.} \emph{et~al.}
\newblock \bibinfo{title}{{Fine structure of neutral and charged excitons in
  self-assembled In(Ga)As/(Al)GaAs quantum dots}}.
\newblock \emph{\bibinfo{journal}{Phys. Rev. B}} \textbf{\bibinfo{volume}{65}},
  \bibinfo{pages}{3216} (\bibinfo{year}{2002}).

\bibitem{Mollow.1969}
\bibinfo{author}{Mollow, B.~R.}
\newblock \bibinfo{title}{{Power Spectrum of Light Scattered by Two-Level
  Systems}}.
\newblock \emph{\bibinfo{journal}{Phys. Rev.}} \textbf{\bibinfo{volume}{188}},
  \bibinfo{pages}{1969--1975} (\bibinfo{year}{1969}).

\bibitem{Davanco.2014.Phys.Rev.B}
\bibinfo{author}{Davan{\c{c}}o, M.}, \bibinfo{author}{Hellberg, C.~S.},
  \bibinfo{author}{Ates, S.}, \bibinfo{author}{Badolato, A.} \&
  \bibinfo{author}{Srinivasan, K.}
\newblock \bibinfo{title}{{Multiple time scale blinking in InAs quantum dot
  single-photon sources}}.
\newblock \emph{\bibinfo{journal}{Phys. Rev. B}} \textbf{\bibinfo{volume}{89}}
  (\bibinfo{year}{2014}).

\bibitem{Fischer.2016}
\bibinfo{author}{Fischer, K.~A.}, \bibinfo{author}{M{\"u}ller, K.},
  \bibinfo{author}{Lagoudakis, K.~G.} \& \bibinfo{author}{Vu{\v{c}}kovi{\'c},
  J.}
\newblock \bibinfo{title}{{Dynamical modeling of pulsed two-photon
  interference}}.
\newblock \emph{\bibinfo{journal}{New J. Phys.}} \textbf{\bibinfo{volume}{18}},
  \bibinfo{pages}{113053} (\bibinfo{year}{2016}).

\bibitem{Fischer.2017}
\bibinfo{author}{Fischer, K.~A.} \emph{et~al.}
\newblock \bibinfo{title}{{Signatures of two-photon pulses from a quantum
  two-level system}}.
\newblock \emph{\bibinfo{journal}{Nat. Phys.}} \textbf{\bibinfo{volume}{13}},
  \bibinfo{pages}{649--654} (\bibinfo{year}{2017}).

\bibitem{Bennett.2008}
\bibinfo{author}{Bennett, A.~J.} \emph{et~al.}
\newblock \bibinfo{title}{{Indistinguishable photons from a diode}}.
\newblock \emph{\bibinfo{journal}{Appl. Phys. Lett.}}
  \textbf{\bibinfo{volume}{92}}, \bibinfo{pages}{193503}
  (\bibinfo{year}{2008}).

\bibitem{Varoutsis.2005.Phys.Rev.B}
\bibinfo{author}{Varoutsis, S.} \emph{et~al.}
\newblock \bibinfo{title}{{Restoration of photon indistinguishability in the
  emission of a semiconductor quantum dot}}.
\newblock \emph{\bibinfo{journal}{Phys. Rev. B}} \textbf{\bibinfo{volume}{72}},
  \bibinfo{pages}{681} (\bibinfo{year}{2005}).

\bibitem{Dada.2016}
\bibinfo{author}{Dada, A.~C.} \emph{et~al.}
\newblock \bibinfo{title}{{Indistinguishable single photons with flexible
  electronic triggering}}.
\newblock \emph{\bibinfo{journal}{Optica}} \textbf{\bibinfo{volume}{3}},
  \bibinfo{pages}{493} (\bibinfo{year}{2016}).

\bibitem{Gazzano.2013}
\bibinfo{author}{Gazzano, O.} \emph{et~al.}
\newblock \bibinfo{title}{{Bright solid-state sources of indistinguishable
  single photons}}.
\newblock \emph{\bibinfo{journal}{Nat. Commun.}} \textbf{\bibinfo{volume}{4}},
  \bibinfo{pages}{1425} (\bibinfo{year}{2013}).

\bibitem{Bockler.2008}
\bibinfo{author}{B{\"o}ckler, C.} \emph{et~al.}
\newblock \bibinfo{title}{{Electrically driven high-Q quantum dot-micropillar
  cavities}}.
\newblock \emph{\bibinfo{journal}{Appl. Phys. Lett.}}
  \textbf{\bibinfo{volume}{92}}, \bibinfo{pages}{091107}
  (\bibinfo{year}{2008}).

\bibitem{Maier.2014.}
\bibinfo{author}{Maier, S.} \emph{et~al.}
\newblock \bibinfo{title}{{Bright single photon source based on self-aligned
  quantum dot-cavity systems}}.
\newblock \emph{\bibinfo{journal}{Opt. Express}} \textbf{\bibinfo{volume}{22}},
  \bibinfo{pages}{8136--8142} (\bibinfo{year}{2014}).

\bibitem{Unsleber.2015}
\bibinfo{author}{Unsleber, S.} \emph{et~al.}
\newblock \bibinfo{title}{{Deterministic generation of bright single resonance
  fluorescence photons from a Purcell-enhanced quantum dot-micropillar
  system}}.
\newblock \emph{\bibinfo{journal}{Opt. Express}} \textbf{\bibinfo{volume}{23}},
  \bibinfo{pages}{32977--32985} (\bibinfo{year}{2015}).

\bibitem{Strau.2016}
\bibinfo{author}{Strau{\ss}, M.} \emph{et~al.}
\newblock \bibinfo{title}{{Photon-statistics excitation spectroscopy of a
  single two-level system}}.
\newblock \emph{\bibinfo{journal}{Phys. Rev. B}} \textbf{\bibinfo{volume}{93}}
  (\bibinfo{year}{2016}).

\bibitem{Nguyen.2012}
\bibinfo{author}{Nguyen, H.~S.} \emph{et~al.}
\newblock \bibinfo{title}{{Optically gated resonant emission of single quantum
  dots}}.
\newblock \emph{\bibinfo{journal}{Phys. Rev. Lett.}}
  \textbf{\bibinfo{volume}{108}}, \bibinfo{pages}{057401}
  (\bibinfo{year}{2012}).

\end{thebibliography}
\end{document}